# Impact of holding umbrella on uni- and bi-directional pedestrian flow: Experiments and modeling

Ning Guo [1], Qing-Yi Hao [2], Rui Jiang [3,*], Mao-Bin Hu [1], Bin Jia [3]

*1 School of Engineering Science, University of Science and Technology of China, Hefei 230026, P. R. China*

*2 School of Mathematics and Computational Science, Anqing Teachers College, Anqing 246133, P. R. China*

*3 MOE Key Laboratory for Urban Transportation Complex Systems Theory and Technology, Beijing Jiaotong University, Beijing 100044, P.R. China*

**Abstract**

In this paper, the impact of holding umbrella on the uni- and bi-directional flow has been investigated via experiment and modeling. In the experiments, pedestrians are required to walk clockwise/anti-clockwise in a ring-shaped corridor under normal situation and holding umbrella situation. No matter in uni- or bi-directional flow, the flow rate under holding umbrella situation decreases comparing with that in normal situation. In bidirectional flow, pedestrians segregate into two opposite moving streams very quickly under normal situation, and clockwise/anti-clockwise walking pedestrians are always in the inner/outer lane due to right-walking preference. Under holding umbrella situation, spontaneous lane formation has also occurred. However, when holding umbrella, pedestrians may separate into more than two lanes. Moreover, the merge of lanes have been observed, and clockwise/anti-clockwise pedestrians are not always in the inner/outer lane. To model the flow dynamics, an improved force-based model has been proposed. The contact force between umbrellas has been taken into account. Simulation results are in agreement with the experimental ones.

## 1. Introduction

Recently, the study of pedestrian dynamics has attracted considerable attention in many research fields (Helbing, 2001; Daamen and Hoogendoorn, 2003). In order to enhance level-of-service and efficiency of pedestrian facilities and improve crowd safety during mass events or evacuation processes, it is necessary to investigate the local interaction laws underlying pedestrian dynamics.

Current pedestrian dynamics studies mainly focus on two fields. One is pedestrian evacuation in emergence situations. Helbing et al (2000a) took into account the panic degree in the social force model, and observed faster-is-slower effect and arching in evacuation. Garcimartín et al. (2014) launched a controlled experiment by allowing pushing one another or not to create competitive or non-competitive egress condition. It was found that competitive egress produces longer evacuation time. Studies showed that exit width and exit distribution also have significant impact on escape efficiency (Moussaïd et al.,

---

* Corresponding author.
Email address: jiangrui@bjtu.edu.cn, humaobin@ustc.edu.cn, bjia@bjtu.edu.cn

2011; Hou et al, 2014).

The other field is pedestrian flow under normal condition. In bidirectional flow, lane formation is a common phenomenon (Zhang et al 2012). In intersection areas, stripes will form allowing crossing flows to penetrate each other (Helbing and Johansson, 2011). Hoogendoorn and Daamen (2005) found that to reduce the interaction between pedestrians, zipper effect occurs in the narrow corridor. As a kind of social animal, the social group also affects the walking behavior (Moussaïd et al, 2010). In particular, when the pedestrian density is large, the turbulent flow and stampede accident (Helbing et al, 2007) potentially happen.

For pedestrian flow under normal condition, the velocity-density diagram and flow-density diagram (fundamental diagram) are used to reveal the basic features of pedestrian dynamics (Seyfried et al, 2005; Chattaraj et al 2009; Suma et al, 2012; Yanagisawa et al, 2012; Flötteröd and Lämmel, 2015), such as pedestrian flow capacity, transition from free flow to congestion, and so on.

The velocity-density diagram and flow-density diagram are affected by various factors. Chattaraj et al (2009) developed experiments with participants from different countries with distinct cultures, finding that the speed of German participants is more sensitive to density than Indian, and the movement of Indians is more effective than Germans. Suma et al (2012) conducted an experiment in which participants manipulate cellular phones, and found that phones result in smaller strength and range of anticipation, weakening the walking speed. In the view-limited condition (Guo et al, 2015), pedestrians prefer a lower speed no matter in uni- or bi- directional flow, comparing to normal condition. Nagai et al (2005) investigated pedestrians going on all fours in counterflow, finding that the speed decreases more abruptly with increasing density. Social groups are common in pedestrian dynamics, and the growing group size reduces the average speed (Moussaïd et al, 2010). Yanagisawa et al (2012) performed an experiment demonstrating that slow rhythm can improve the flow rate in a congested situation.

Lane formation is viewed as segregation phenomenon that oppositely moving pedestrians emerge into lanes of uniform walking direction. It is a self-organized collaborative pattern of motion originating from simple pedestrian interactions (Navin and Wheeler, 1969; Helbing et al, 2000b; Schadschneider et al, 2003; Zhang and Seyfried, 2014). Helbing and Johansson (2011) found that in order to minimize frictional effects and energy consumption, pedestrians follow others in the same direction. In simulation, even if right/left-walking preference is not considered, lane formation also occurs. In experiments by Kretz et al (2006), when two lanes are formed, pedestrians are always on the right hand side. Guo et al (2012) confirmed the look-ahead behavior of pedestrians in bidirectional flow by experiments. Nowak and Schadschneider (2012) introduced anticipation mechanism into the floor field model, which reproduces lane formation at high density. Moussaïd et al (2012) observed that the lanes are instable, and inferred that velocity difference lead to the phenomenon in the cognitive model. In the experiments of Zhang et al (2012), with instruction requiring pedestrians to use specified exits, lane construction keeps varying. Without instruction, lanes remain stable form in the whole process. However, the mechanism causing lane formation is still not completely interpreted.

In the rainy/snowy day or sun-shining day in hot summer, the pedestrians walking outdoor usually hold umbrellas. However, few studies investigate the influence of holding umbrella on pedestrian dynamics. In this paper, the impact of holding umbrella on pedestrian flow is explored. The motivation of the study is twofold. Firstly, we concern about the quantitative effect of holding umbrella on pedestrian flow fundamental diagram. Secondly, we examine the impact of holding umbrella on the lane formation in bidirectional pedestrian flow. The experiments showed that under normal condition, pedestrians walking in different directions always separate into two stable lanes very quickly. However, when holding umbrella, pedestrians may separate into more than two lanes. Moreover, the merge of lanes have been observed. To model the pedestrian flow, an improved force-based model has been proposed. Simulation results are in agreement with the experiment.

The remainder of this paper is organized as follows. Section 2 describes the experimental setup. Section 3 presents the experimental results. Section 4 proposes improved force-based model and presents the simulation results. Section 5 concludes the paper.

## 2. Experimental setup

We performed the experiments in the afternoon on December 13, 2014 at Anqing Teachers College. As in previous experiment carried out in the morning on the same day (Guo et al, 2015), an artificial ring-shaped corridor was utilized to conduct experiments. The inner radius and external radius of the corridor were 2 m and 5 m respectively, and the boundaries were marked with pasters (Fig. 1). The area of the experimental corridor is thus approximately 66 $m^2$. 100 participants (undergraduate, 25 male, 75 female), which were different from participants in the morning experiment, participated in the experiments, and everyone had a serial number from 1 to 100. Participants with odd/even number held blue/orange umbrellas. The diameter of the umbrella was 1 m. The participants were naïve to the purpose of experiments.

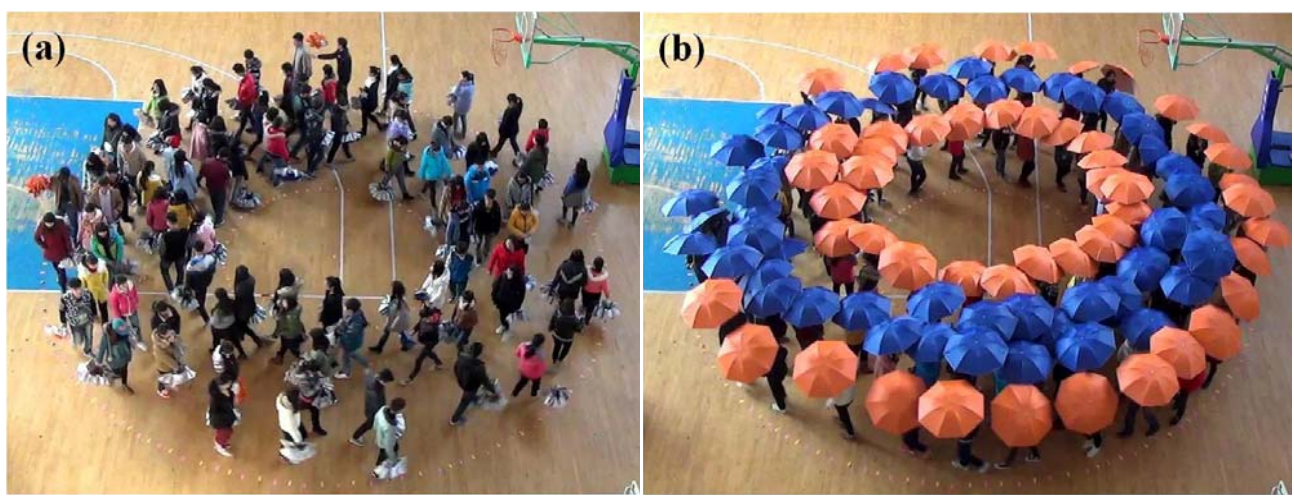

Fig.1 Snapshots of the experiments. (a) normal situation, N=90, t=70s. (b) umbrella situation, N=100, t=90s.

Seven sets of experiments have been implemented. Chronologically the participant number in each set was N=100, 60, 40, 80, 70, 90, 50. Participant numbers were 1-100 in set 1, 1-60 in set 2, 61-100 in set 3, 1-80 in set 4, and 31-100 in set 5, 1-90 in set 6, 50-100 in set 7.

For the last two sets of experiments, firstly the participants were requested to walk without holding umbrellas (normal situation, see Fig.1a). Initially, all the participants were distributed randomly in the corridor. They were requested to walk in the same direction. After some time, they were asked to stop. Then N/2 participants with odd/even numbers were asked to walk clockwise, and the other N/2 with even/odd numbers walked anti-clockwise.

Next these participants were requested to hold up the umbrellas (umbrella situation, see Fig.1b). As under normal situation, the participants were firstly requested to walk in the same direction. Then N/2 participants were asked to change their walk direction. This ends one set of experiment.

For the first five sets of experiments, participants were only requested to hold up the umbrellas. They were asked to firstly walk uni-directionally and then bi-directionally.

The experimental processes were recorded by video camera (SONY HDR-CX510E). The uni- and bi-directional flow rates were extracted manually.

## 3. Experimental results

*3.1 Experimental results under normal situation*

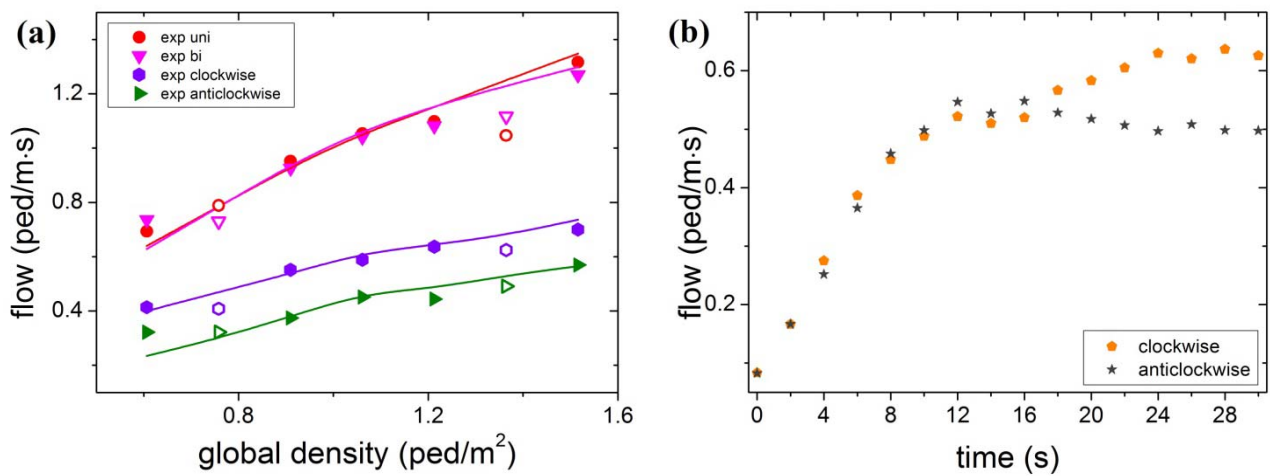

Fig. 2 (a) The fundamental diagram under normal situation. Scattered data are experimental results and solid lines are simulation results. The abbreviations in the legend are as follows in this figure and following figures. uni: unidirectional flow rate; bi: bidirectional flow rate; clockwise/anticlockwise: flow rate of pedestrians walking in clockwise/anticlockwise direction in bidirectional flow. (b) The evolution of clockwise and anti-clockwise flow rates over time in the experiment under normal situation, N=90.

Fig. 2a presents the fundamental diagrams of uni- and bi-directional flow under normal view condition. The spaced symbols are results of present experiments and the filled symbols are results of previous morning experiments (Guo et al, 2015).

The flow rate is defined as the number of pedestrians passing by a location in the statistical time/(statistical time × corridor width). Unless otherwise mentioned, the statistical time is experimental time minus initial 15 seconds so that the transient time is excluded. Although the participants in the morning experiments are different from that in the afternoon experiments, the flow rates are essentially consistent. The flow rates increase with the global density and the decreasing branch of fundamental diagram is not observed in the experiments since the pedestrian number in experiments is not enough: the maximum pedestrian number 100 only corresponds to the global density 1.51 ped/m$^2$.

The experimental results of bidirectional flow in the afternoon experiment are also similar to that in the morning experiment: two lanes spontaneously form fast, and the clockwise/anti-clockwise walking pedestrians are always in the inner/outer lane due to right walking preference. Fig.2b shows the typical evolution process of clockwise and anti-clockwise flow rates over time. Initially the flow rates of both directions are approximately equal to each other. However, after about 18 seconds, the clockwise flow rate begins to exceed the anti-clockwise flow rate, which indicates the lane has formed.

*3.2 Impact of holding umbrella*

The fundamental diagram of unidirectional flow in which participants hold umbrella is presented in Fig. 3a. The flow rate increases until the global density reaches 1.21 ped/m$^2$ (N=80), and then it becomes basically saturated. Quantitatively, the flow rate is about 0.1 ped/m·s smaller than that under normal situation when N≤80. Then the flow rate difference increases with the further increase of N[1]. At the global density 1.51 ped/m$^2$, the flow rate difference reaches about 0.3 ped/m·s, see Fig.3b. This is easy to understand since umbrellas increase the interaction distance between pedestrians.

Now we show the experimental results of bidirectional flow in the 6th set of experiment, in which pedestrian number N=90. Spontaneous lane formation phenomenon has also occurred, as Fig.4a and Supplemental Video S1 show. Comparing to the normal situation, three lanes have formed. Pedestrians walking in clockwise direction segregate into two lanes, and they move along the outer and inner ring, respectively. Pedestrians walking in anti-clockwise direction form only one lane in the middle. In the outer lane, the pedestrians walk in platoon. The leading pedestrian is very uncomfortable due to frequent interactions with opposite moving ones. At about 55 second, a gap appears in the middle lane. The leading pedestrian in the outer lane thus crosses the gap and goes into the inner lane to avoid the frequent interactions. The following pedestrians in the outer lane follow the leading pedestrian. As a result, two clockwise direction lanes merge together, see Fig.4b and Supplemental Video S1. The two lanes have been stable throughout the experiment, see Fig.4c. Fig.4d shows that when two clockwise direction lanes merge, the difference between clockwise and anti-clockwise flow rates increases.

---

[1] Since the set of experiment N=90 is carried out late, the participants are tired. Therefore, the flow rate difference is smaller than expected.

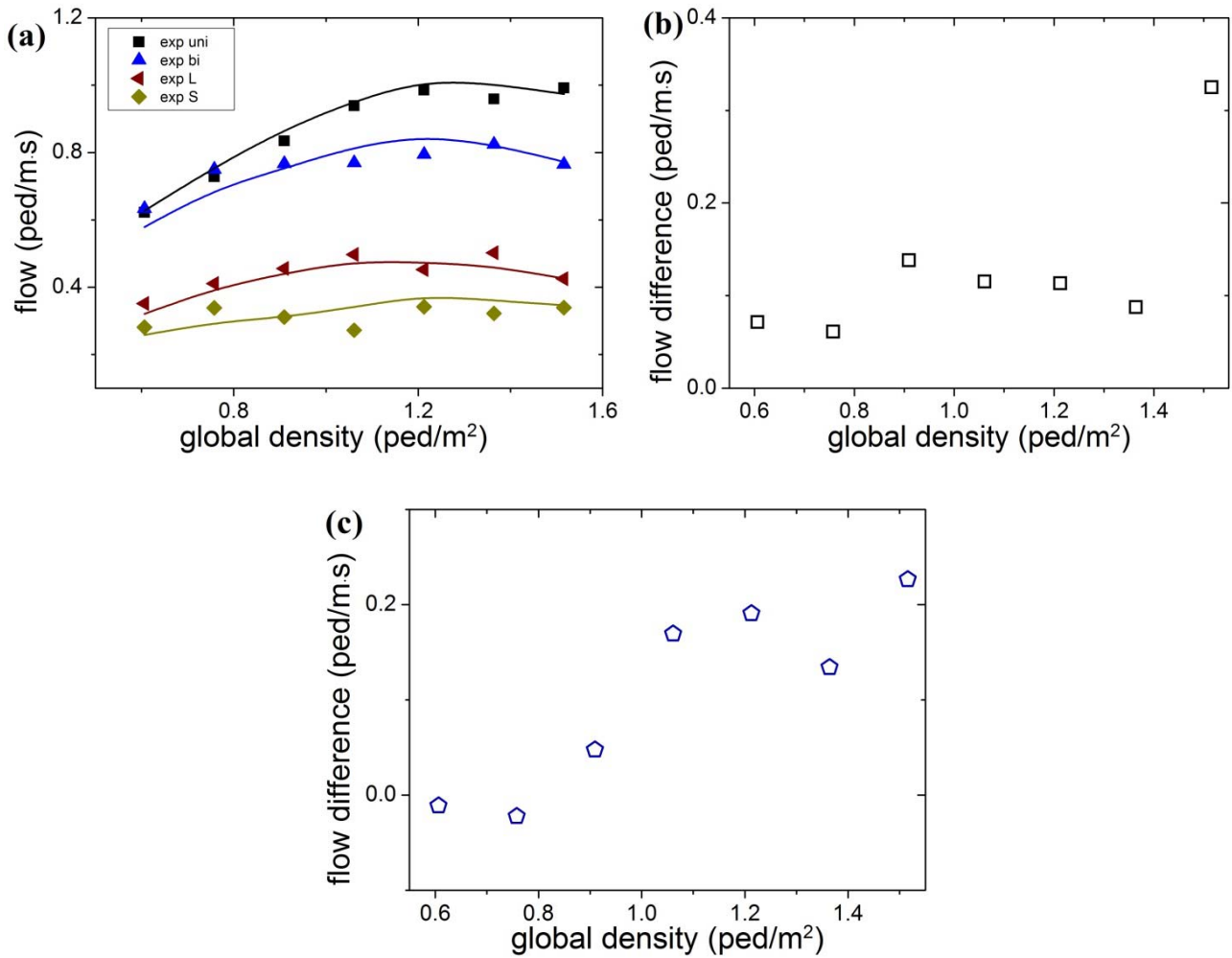

Fig. 3 (a) The fundamental diagram under holding umbrella situation. Scattered data are experimental results and solid lines are simulation results. L/S: the larger/smaller one between the clockwise flow rate and the anticlockwise flow rate. (b) Flow difference between unidirectional experiments under normal situation and holding umbrella situation. (c) Flow difference between uni- and bi-directional experiments under holding umbrella situation.

Next we present the experimental results of bidirectional flow in the 1st set of experiment, in which pedestrian number N=100. Three lanes have also formed, see Fig.1b. The leading pedestrian in the outer lane is also very uncomfortable. However, since no gap has appeared in the middle lane, the leading pedestrian has no chance to merge into the inner lane. The three lanes have been maintained throughout the set of experiment.

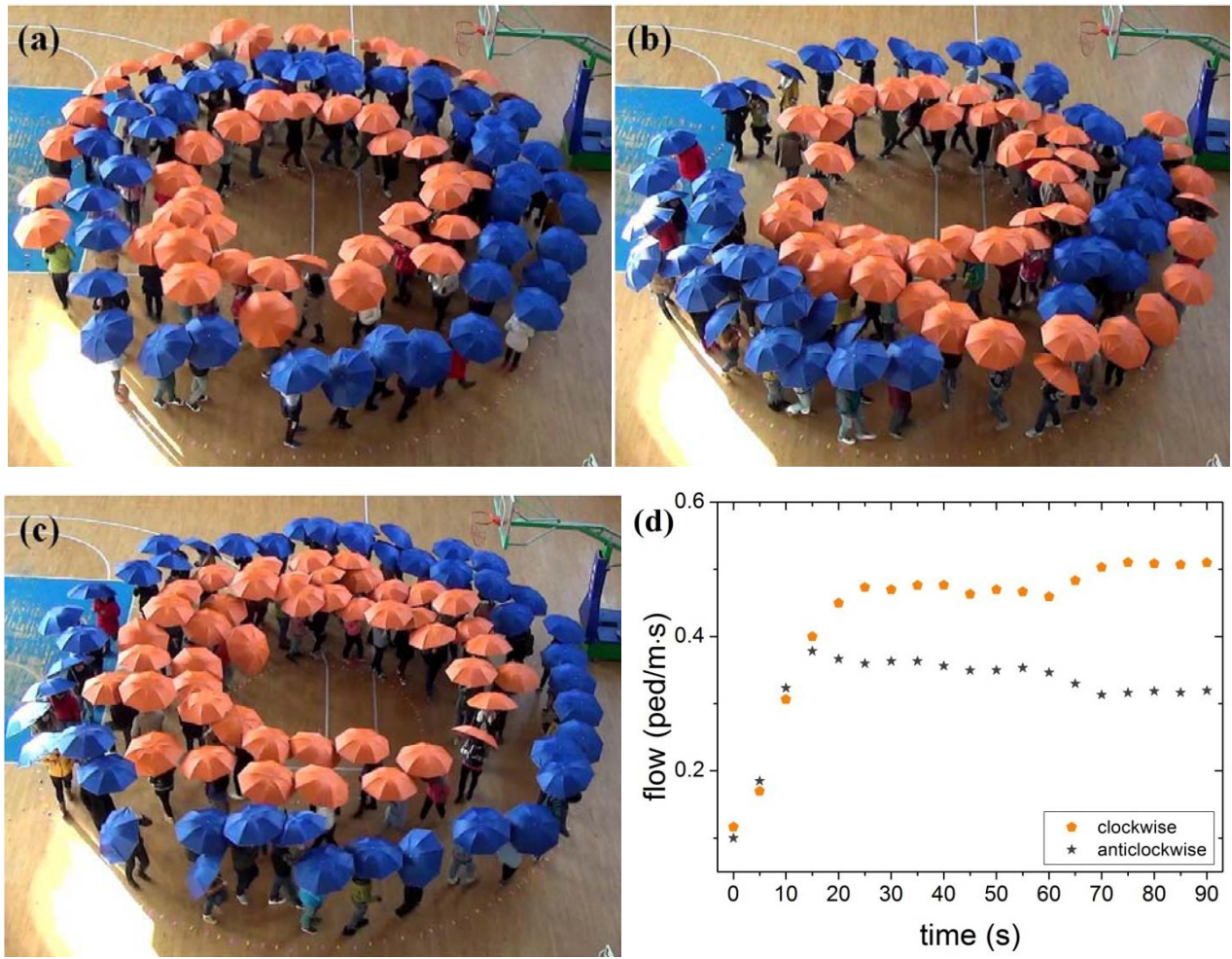

Fig. 4 Snapshots of the experiment N=90 under holding umbrella situation. (a) t=35s. (b) t=60s. (c) t=105s. (d) The evolution of clockwise and anti-clockwise flow rates over time.

In the experimental results of bidirectional flow in the 7th set of experiment, in which pedestrian number N=50, pedestrians quickly form four lanes, see Fig.5a. At about 32 second, two clockwise direction lanes and two anticlockwise direction lanes merge together respectively, see Fig.5b and Supplemental Video S2. Note that pedestrians in anti-clockwise direction walk in the inner ring (Fig.5c).

Similar phenomenon is also observed in experiments of N=70, 40, see Table 1. Comparing to the normal situation, the clockwise walking pedestrians are not always in the inner lane under the holding umbrella situation. This might be because umbrellas reduce the free space among pedestrians, which suppresses the right-walking preference of pedestrians.

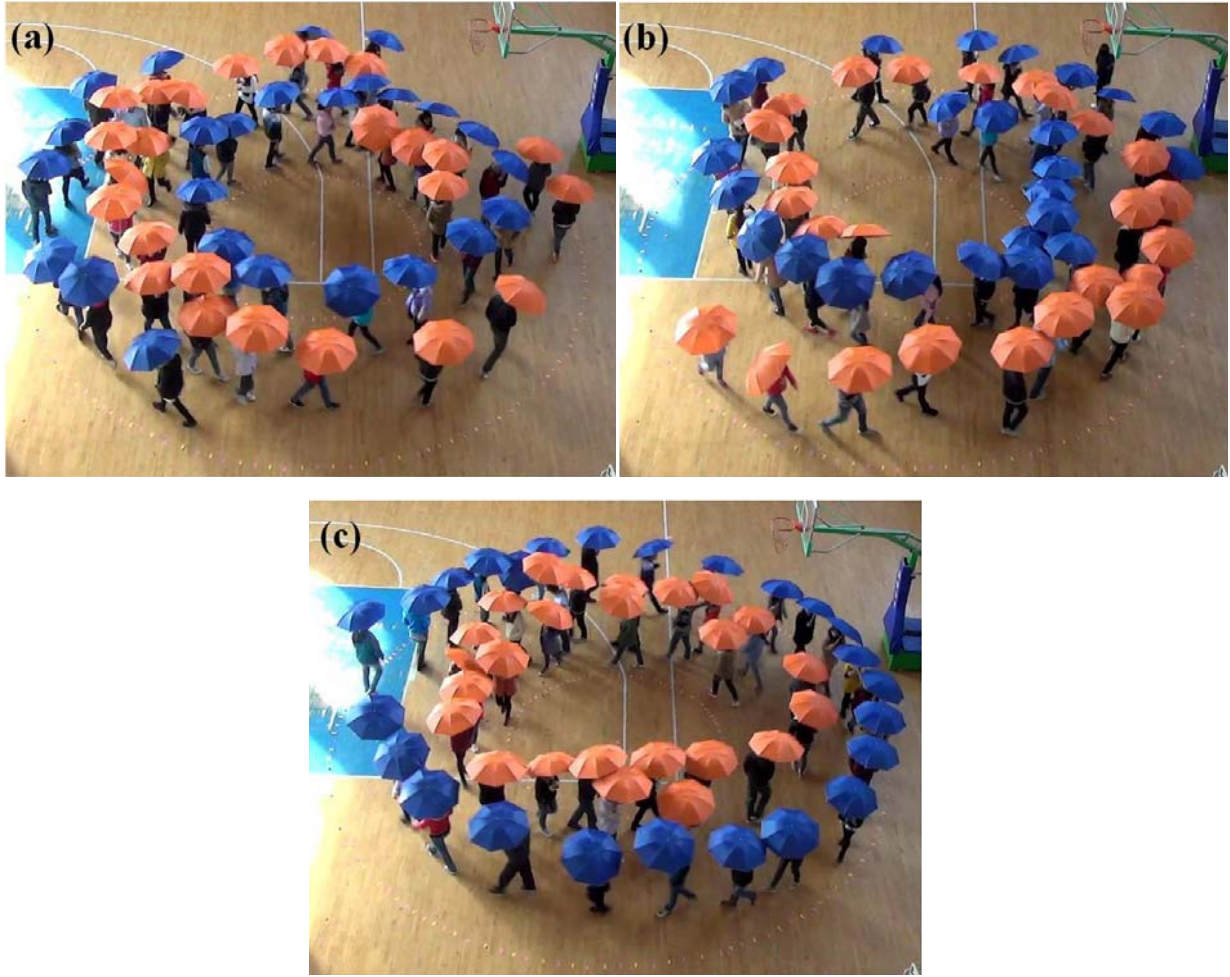

Fig. 5 Snapshots of the experiment N=50 under holding umbrella situation. (a) t=25s. (b) t=35s. (c) t=45s.

Table 1 lane number and walking direction of inner lane under holding umbrella situation

| Pedestrian number | Lanes number | Direction of inner lane |
|---|---|---|
| 40 | 3 to 2 | anticlockwise |
| 50 | 4 to 2 | anticlockwise |
| 60 | 2 | clockwise |
| 70 | 3 to 2 | anticlockwise |
| 80 | 3 | anticlockwise |
| 90 | 3 to 2 | clockwise |
| 100 | 3 | clockwise |

The flow rate of bidirectional flow is also presented in Fig.3a[2]. The flow rate increases until the global density reaches 0.75 ped/m$^2$ (N=50), and then it becomes basically saturated. Different from that under normal situation, in which bidirectional flow rate is essentially equal to the unidirectional flow rate (Fig.2a), the bidirectional flow rate under umbrella situation is smaller than the unidirectional flow rate by 0.2 ped/m·s, when the global density is larger than 1.05 ped/m$^2$ (N=70), see Fig.3c. This is because even if pedestrians in opposite directions have separated, the interactions between umbrellas in opposite directions are more intense and frequent than the physical body interactions under normal situation.

---

[2] The flow rate is calculated in the stationary state after lanes merge, i.e., lane number does not change.

## 4. Modeling and simulation

*4.1 Model*

In our previous paper, we have proposed a force-based model to simulate the unidirectional and bidirectional pedestrian flow under normal condition and view-limited condition (Guo et al, 2015). In this paper, we generalize the model to describe the impact of holding umbrella. For self-consistency of the paper, we firstly review the model (Guo et al, 2015) in this subsection.

In the force-based model, the motion of a pedestrian is described by three different components: driving force, repulsive force between pedestrians, and repulsive force between pedestrian and wall. The equation of motion is as follows,

$$m_i \frac{d\vec{v_i}}{dt} = \vec{f_i^0} + \sum_j \vec{f_{ij}} + \sum_W \vec{f_i^{wall}} \quad (1)$$

Here $m_i$ is mass of pedestrian $i$. The driving force $\vec{f_i^0}$ represents the pedestrian's motivation to walk in a given direction with desired speed, which reads

$$\vec{f_i^0} = m_i \frac{v_i^0 \vec{e_i^0} - \vec{v_i(t)}}{\tau} \quad (2)$$

Here $v_i^0$ and $\vec{e_i^0}$ are respectively magnitude and direction of the pedestrian's desired speed. $\vec{v_i(t)}$ is the current speed, and $\tau$ denotes the relaxation time.

The direction of the desired speed is formulated as follows,

$$v_i^0 \vec{e_i^0} = \begin{cases} v_i^0 \cdot h(i) \cdot t(i) \cdot \left(-\frac{y_{i_c} - y_i}{d_{min}}, \frac{x_{i_c} - x_i}{d_{min}}\right), & \delta = 0 \,\&\, d_{min} < D_{cri1} \quad (3a) \\ 0, & \delta = 1 \,\&\, d_{min} < D_{cri2} \quad (3b) \\ v_i^0 \cdot \left(\cos(H_i^o), \sin(H_i^o)\right), & \text{otherwise} \quad (3c) \end{cases}$$

$$t(i) = \begin{cases} 1, & D_i \geq D_{i_c} \\ -1, & D_i < D_{i_c} \end{cases} \quad (4)$$

$$h(i) = \begin{cases} 1, & clockwise \\ -1, & anticlockwise \end{cases} \quad (5)$$

Here $x_i$, $x_{i_c}$ and $y_i$, $y_{i_c}$ are respectively the horizontal and vertical coordinates of the center of pedestrian $i$ and pedestrian $i_c$. $i_c$ denotes the closest pedestrian to pedestrian $i$ among the ones that locate in the view angle range of $\alpha$ to pedestrian $i$. $d_{min}$ means the distance between pedestrian $i$ and pedestrian $i_c$. $D_{cri1}$ and $D_{cri2}$ are two threshold parameters. $\delta = 0/1$ means that pedestrians $i$ and $i_c$ walk in opposite/same direction. Eq.(3a) means that if pedestrians $i$ and $i_c$ walk in opposite direction and their distance is smaller than $D_{cri1}$, the desired walking direction of pedestrian *i* changes to the vertical direction of the line between the two pedestrians, see Fig.8 in Guo et al (2015). $D_i(D_{i_c})$ is distance between pedestrian *i* ($i_c$) and the circle center. Thus $t(i)$ is used to determine the right/left drift rules: comparing to pedestrian $i_c$, if pedestrian *i* is more far away from circle center, he/she will move to the outside, and vice versa. We would like to mention that under this circumstance, to mimic the behavior that a pedestrian walks diagonally without turning his/her body, we assume that the view direction keeps unchanged, which is no longer the same as the desired walking direction, see Fig.8 and Appendix B in Guo et al (2015).

Eq.(3b) means that if pedestrians *i* and $i_c$ walk in the same direction and their distance is smaller than $D_{cri2}$, pedestrian *i* prefers to stop so that his/her desired speed becomes zero.

Under other circumstances, pedestrians walk along the ring-shaped corridor, see Eq.(3c). $H_i^o$ is the angle of the moving direction

$$H_i^o = H_i^s + H_i^D + \varepsilon(t) + \theta \tag{6}$$

The components are explained below:

➤ $H_i^s$ is the tangent direction of the ring

$$\left(\cos(H_i^s), \sin(H_i^s)\right) = h(i) \cdot \left(\frac{y_i - y_o}{D_i}, -\frac{x_i - x_o}{D_i}\right) \tag{7}$$

where $x_o$ and $y_o$ are respectively the horizontal and vertical coordinates of the center of corridor.

➤ $H_i^D$ is a drift angle toward the corridor center

$$H_i^D = h(i) \cdot \eta \tag{8}$$

$\eta$ is the magnitude of the drift angle. This drift angle is introduced because pedestrians walk in an annulus corridor. Otherwise, a pedestrian always gradually approaches the outer boundary.

➤ $\varepsilon(t)$ is the deviation angle in the current time

$$\varepsilon(t) = \begin{cases} \xi, & \delta = 0 \ \& \ d_{\min}(t) > D_{cri1} \ \& \ d_{\min}(t - \Delta t) < D_{cri1} & (9a) \\ \xi, & \delta = 1 \ \& \ d_{\min}(t) > D_{cri2} \ \& \ d_{\min}(t - \Delta t) < D_{cri2} & (9b) \\ \varepsilon(t - \Delta t), & \text{otherwise} & (9c) \end{cases}$$

Here $\xi$ is a random angle value in the interval $[-\varphi, \varphi]$. $t - \Delta t$ indicates the last time step. Eqs.(9a) and (9b) mean that the closest pedestrian has stepped out of the threshold distance, so that a pedestrian needs to re-choose the desired direction. Otherwise, the desired direction remains unchanged.

➤ $\theta$ is the right-walking preference angle, which denotes the extent of right-walking preference.

$$\left(\cos(H_i^o), \sin(H_i^o)\right) = h(i) \cdot \left(\frac{y_i - y_o}{D_i} \cdot \cos\left(H_i^D + \varepsilon(t) + \theta\right) + \frac{x_i - x_o}{D_i} \cdot \sin\left(H_i^D + \varepsilon(t) + \theta\right),\right.$$

$$\left. -\frac{x_i - x_o}{D_i} \cdot \cos\left(H_i^D + \varepsilon(t) + \theta\right) + \frac{y_i - y_o}{D_i} \cdot \sin\left(H_i^D + \varepsilon(t) + \theta\right)\right) \tag{10}$$

The repulsive force $\vec{f}_{ij}$ describes interactions between pedestrian $i$ and other pedestrians, which includes socio-psychological force to stay away from others and physical contact force. The force is slightly modified compared with that in social force model (Helbing et al, 2000a).

$$\vec{f}_{ij} = \begin{cases} A \cdot \exp[(r_{ij} - d_{ij})/B]\vec{n}_{ij} + g(r_{ij} - d_{ij})(k\vec{n}_{ij} + \kappa \Delta v^t_{ji}\vec{t}_{ij}), & d_{ij} - r_{ij} < D_v \ \& \ \vec{n}_{ij} \cdot \vec{e}^0_i < 0 \quad (11a) \\ g(r_{ij} - d_{ij})(k\vec{n}_{ij} + \kappa \Delta v^t_{ji}\vec{t}_{ij}), & \vec{n}_{ij} \cdot \vec{e}^0_i \geq 0 \quad (11b) \\ 0 & \text{otherwise} \quad (11c) \end{cases}$$

$$g(x) = \begin{cases} x, & x > 0 \\ 0, & x \leq 0 \end{cases} \quad (12)$$

In Eq. (11a), the first term on the right hand side denotes the socio-psychological force, and the second term denotes the physical contact force. Pedestrians $i$ and $j$ are regarded as cylinders with radius $r_i$ and $r_j$, respectively. $r_{ij}$ denotes the sum of the radius of the two pedestrians, namely $r_{ij} = r_i + r_j$. $d_{ij}$ is the distance between the centers of the two pedestrians. $D_v$ is the vision distance. $\vec{n}_{ij}$ represents the unit normal vector from pedestrian $j$ to $i$, $\vec{t}_{ij}$ means the unit tangential vector,

$$\vec{n}_{ij} = (\frac{x_i - x_j}{d_{ij}}, \frac{y_i - y_j}{d_{ij}}) \quad (13)$$

$$\vec{t}_{ij} = (-\frac{y_i - y_j}{d_{ij}}, \frac{x_i - x_j}{d_{ij}}) \quad (14)$$

$\Delta v^t_{ji}$ means the tangential velocity,

$$\Delta v^t_{ji} = (\vec{v}_j - \vec{v}_i) \cdot \vec{t}_{ij} \quad (15)$$

where $k\vec{n}_{ij}$ and $\kappa \Delta v^t_{ji}\vec{t}_{ij}$ are body pressure and friction force respectively. $\vec{n}_{ij} \cdot \vec{e}^0_i < 0$ means that pedestrian $j$ is located in the front of the pedestrian $i$. In this case, we assume there is socio-psychological force from pedestrian $j$ on pedestrian $i$, see Eq.(11a). In contrast, when pedestrian $j$ is located behind the pedestrian $i$, we assume there is no socio-psychological force from pedestrian $j$ on pedestrian $i$, see Eq.(11b). Here $A$, $B$, $k$ and $\kappa$ are four parameters.

The repulsive force $\vec{f}^{wall}_i$ from the wall is modeled analogously,

$$\vec{f}^{wall}_i = \begin{cases} A \cdot \exp[(r_i - d^{wall}_i)/B]\vec{n}^{wall}_i + g(r_i - d^{wall}_i)(k\vec{n}^{wall}_i + \kappa \Delta v^t_i \vec{t}^{wall}_i), & d^{wall}_i - r_i < D_v \\ 0, & \text{otherwise} \end{cases} \quad (16)$$

Here, $d^{wall}_i$ means vertical distance to the wall, $\vec{n}^{wall}_i$ denotes the unit normal vector from the wall to pedestrian $i$, and $\vec{t}^{wall}_i$ is the unit vector tangential to the wall. Since speed of the wall is zero, the tangential velocity $\Delta v^t_i = -\vec{v}_i \cdot \vec{t}^{wall}_i$.

Next we generalize the model to consider the impact of holding umbrella. Under the holding umbrella situation, the contact between umbrellas will generate physical force, which should be added into the repulsive force. The repulsive force $\vec{f}_{ij}$ is modified as,

$$\overrightarrow{f_{ij}} = \begin{cases} A \cdot \exp[(r_{ij} - d_{ij})/B]\overrightarrow{n_{ij}} + g(r_{ij} - d_{ij})(k\overrightarrow{n_{ij}} + \kappa \Delta v^t_{ji}\overrightarrow{t_{ij}}) + g(R_{ij} - d_{ij})(k_2\overrightarrow{n_{ij}} + \kappa_2 \Delta v^t_{ji}\overrightarrow{t_{ij}}), & d_{ij} - r_{ij} < D_v \ \& \ \overrightarrow{n_{ij}} \cdot \overrightarrow{e^0_i} < 0 & (17a) \\ g(r_{ij} - d_{ij})(k\overrightarrow{n_{ij}} + \kappa \Delta v^t_{ji}\overrightarrow{t_{ij}}) + g(R_{ij} - d_{ij})(k_2\overrightarrow{n_{ij}} + \kappa_2 \Delta v^t_{ji}\overrightarrow{t_{ij}}), & \overrightarrow{n_{ij}} \cdot \overrightarrow{e^0_i} \geq 0 & (17b) \\ 0 & \text{otherwise} & (17c) \end{cases}$$

In Eq. (17a), the first term on the right hand side denotes the umbrella contact force. Pedestrians $i$ and $j$ hold umbrellas with radius $R_i$ and $R_j$, respectively. $R_{ij}$ denotes the sum of the radius of the two umbrellas, namely $R_{ij} = R_i + R_j$. $k_2$ and $\kappa_2$ are two parameters.

Since the corridor is marked by pasters, the physical contact force between umbrella and wall is ignored. The repulsive force $\overrightarrow{f_i^{wall}}$ from the wall is unchanged.

*4.2 Simulation results under holding umbrella situation*

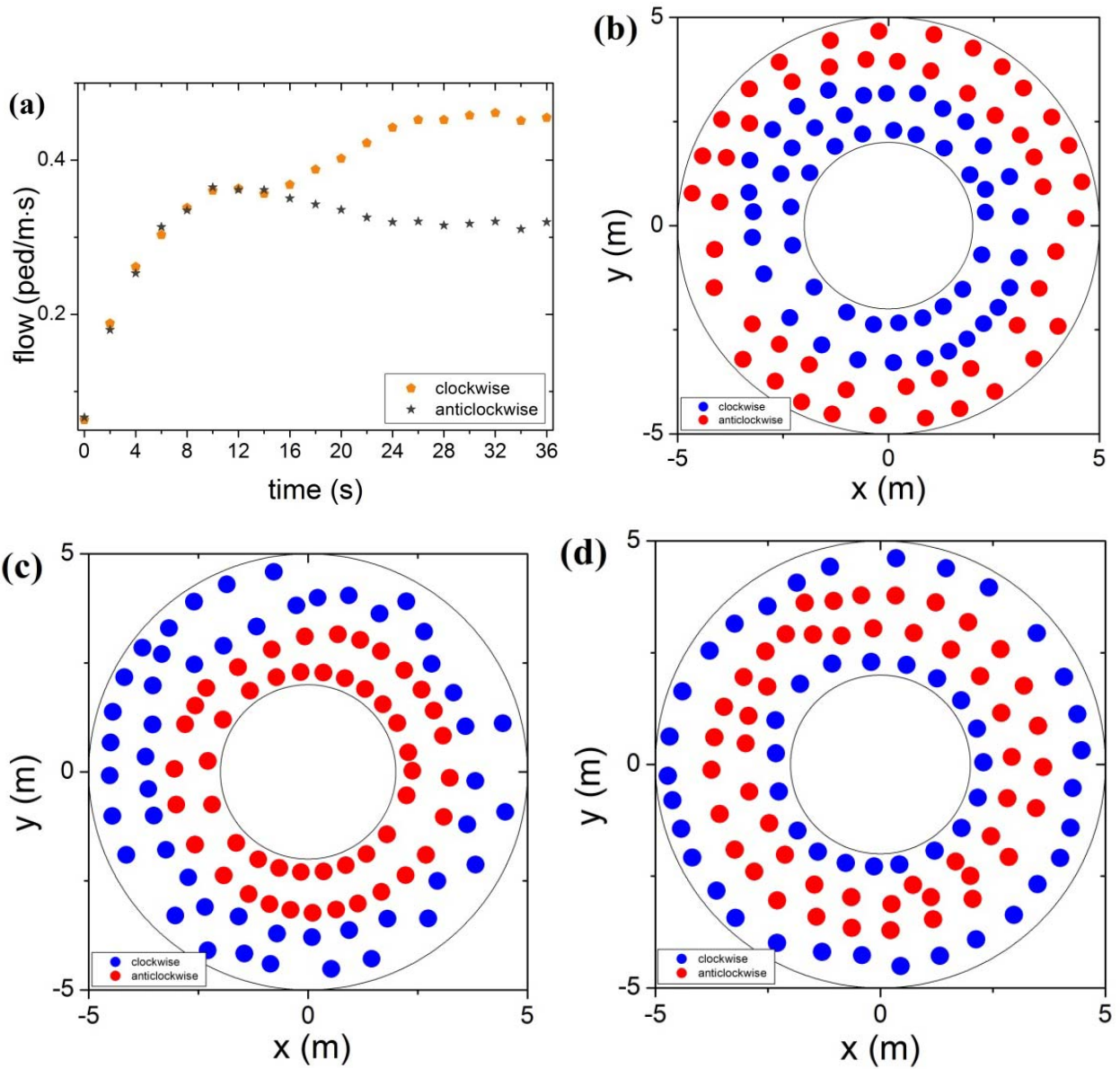

Fig. 6 Simulation results of (a) the evolution of clockwise and anti-clockwise flow rates over time corresponding to (b). (b) (c) and (d) three typical pedestrian configuration, N=100.

As proposed in Sivers et al (2015), the calibration is performed by simulating several runs with different parameters. The parameter values sufficing to achieve a good quantitative fit to the fundamental diagram are selected. Under the holding umbrella situation, the parameter values are: $\tau = 0.5s$, $r_i = 0.2m$, $R_i$=0.5m, $D_v$=5m, $D_{cri1}$=1.2m, $D_{cri2}$=0.775m, α=π/3, $\eta$ =-9π/200, $\varphi$=0, $\Delta t$ = 0.01s, $A$=70N, $B$=0.08m, $k = 2800 kg \cdot s^{-2}$, $\kappa = 700 kg \cdot m^{-1} s^{-1}$, $k_2 = 70 kg \cdot s^{-2}$, $\kappa_2 = 140 kg \cdot m^{-1} s^{-1}$. $m_i$ follows normal distribution with mean of 70kg and standard deviation 8kg, $v_i^0$ follows normal distribution with mean of 1.1 m/s and standard deviation 0.05 m/s. In unidirectional flow, we set $\theta = 0$. For bidirectional flow, the parameter $\theta = -\pi/20$. The number of clockwise walking pedestrians equals to that of anticlockwise walking pedestrians. The simulation results under normal situation are shown in Fig.2a, Tables 2 and 3, see Guo et al. (2015) for more details.

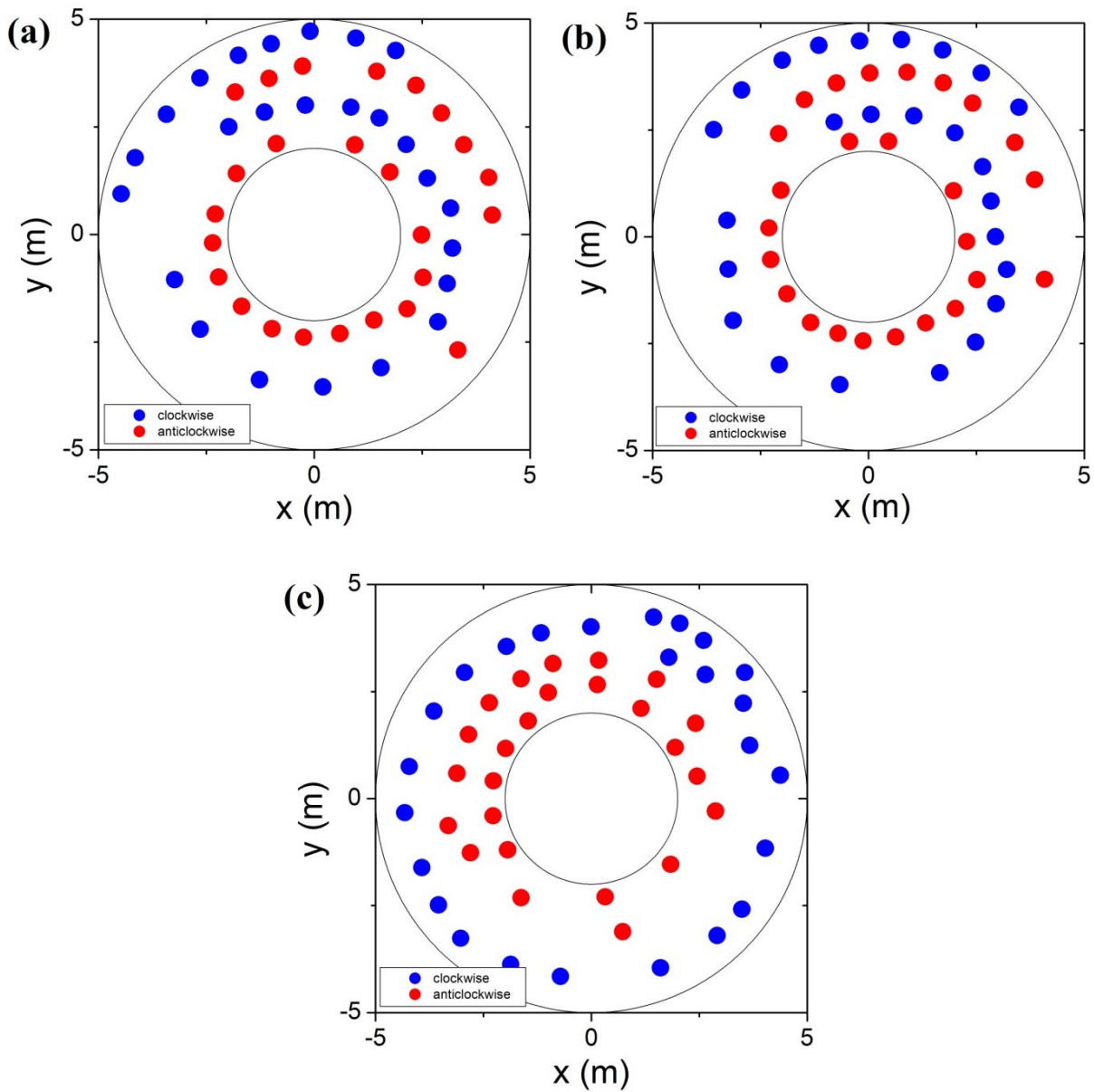

Fig.7 lane merge process, N=50. (a) t=30s, (b) t=35s, (c) t=60s.

Under holding umbrella situation, simulations show that spontaneous lane formation occurs (Fig.6a). Fig.6b-d show three typical patterns. The clockwise/anti-clockwise pedestrians are not always in inner/outer lane (Fig.6c). Moreover, there might form more than two lanes (Fig.6d). Fig.7 shows the transform process that 4 lanes merge into 2 lanes. The pedestrians in middle lane in the anticlockwise direction cross the gap and merge into the inner lane. After that, two lanes keep stable. Fig. 8 presents another lane merge process when the pedestrian number N=80.

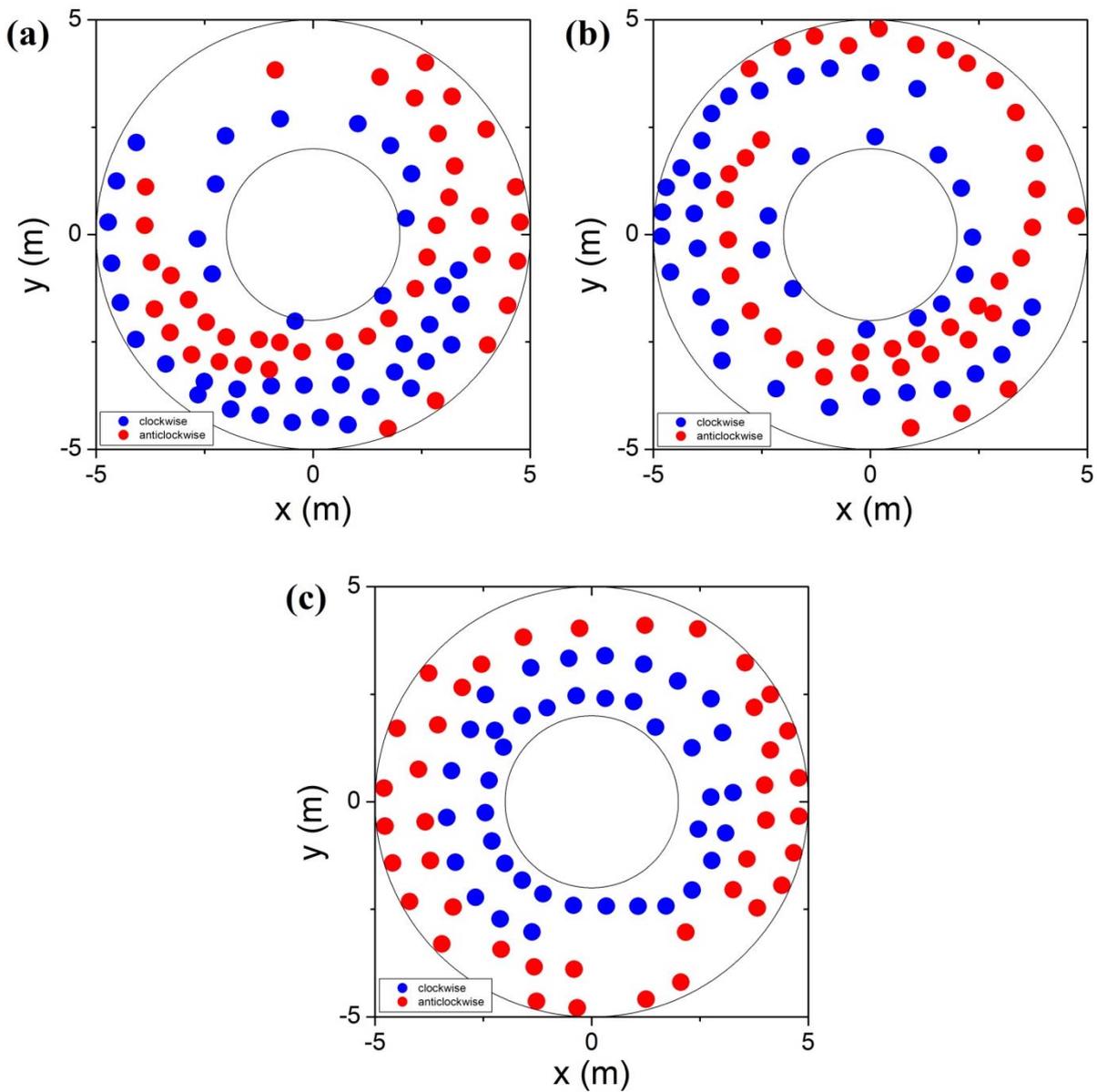

Fig.8 lane merge process, N=80. (a) t=60s, (b) t=70s, (c) t=100s.

Table 2 A quantitative comparison between simulation results and experimental ones of unidirectional flow rate. (unit: ped/m·s)

|  | Normal situation | | | Holding umbrella situation | | |
|---|---|---|---|---|---|---|
|  | Experiment | Simulation | Difference | Experiment | Simulation | Difference |
| 1st set | 1.317 | 1.348 | -0.031 | 0.992 | 0.974 | 0.018 |
| 2nd set | 0.952 | 0.931 | 0.021 | 0.834 | 0.869 | -0.035 |
| 3rd set | 0.694 | 0.640 | 0.054 | 0.622 | 0.627 | -0.005 |
| 4th set | 1.098 | 1.151 | -0.053 | 0.985 | 1.014 | -0.029 |
| 5th set | 1.054 | 1.053 | 0.001 | 0.939 | 0.953 | -0.014 |
| 6th set | 1.066 | 1.230 | -0.164 | 0.959 | 1.004 | -0.045 |
| 7th set | 0.789 | 0.785 | 0.004 | 0.728 | 0.756 | -0.028 |
| RMSE |  |  | 0.0697 |  |  | 0.0278 |

Table 3 A quantitative comparison between simulation results and experimental ones of bidirectional flow rate under normal situation. (unit: ped/m·s)

|  | Bidirectional flow | | | Clockwise flow | | | Anticlockwise flow | | |
| --- | --- | --- | --- | --- | --- | --- | --- | --- | --- |
|  | Experiment | Simulation | Difference | Experiment | Simulation | Difference | Experiment | Simulation | Difference |
| 1st set | 1.269 | 1.298 | -0.029 | 0.700 | 0.733 | -0.033 | 0.569 | 0.565 | 0.004 |
| 2nd set | 0.926 | 0.939 | -0.013 | 0.552 | 0.549 | 0.003 | 0.374 | 0.390 | -0.016 |
| 3rd set | 0.736 | 0.629 | 0.107 | 0.414 | 0.394 | 0.020 | 0.322 | 0.235 | 0.087 |
| 4th set | 1.081 | 1.145 | -0.064 | 0.637 | 0.655 | -0.018 | 0.444 | 0.490 | -0.046 |
| 5th set | 1.041 | 1.064 | -0.023 | 0.589 | 0.600 | -0.011 | 0.452 | 0.464 | -0.012 |
| 6th set | 1.117 | 1.230 | -0.113 | 0.625 | 0.690 | -0.065 | 0.492 | 0.540 | -0.048 |
| 7th set | 0.731 | 0.783 | -0.052 | 0.408 | 0.478 | -0.070 | 0.323 | 0.305 | 0.018 |
| RMSE |  |  | 0.0682 |  |  | 0.0398 |  |  | 0.0426 |

Table 4 A quantitative comparison between simulation results and experimental ones of bidirectional flow rate under holding umbrella situation. (unit: ped/m·s)

|  | Bidirectional flow | | | Flow of one direction | | | Flow of other direction | | |
| --- | --- | --- | --- | --- | --- | --- | --- | --- | --- |
|  | Experiment | Simulation | Difference | Experiment | Simulation | Difference | Experiment | Simulation | Difference |
| 1st set | 0.765 | 0.772 | -0.007 | 0.425 | 0.426 | -0.001 | 0.340 | 0.346 | -0.006 |
| 2nd set | 0.767 | 0.752 | 0.015 | 0.456 | 0.442 | 0.014 | 0.311 | 0.310 | 0.001 |
| 3rd set | 0.633 | 0.579 | 0.054 | 0.352 | 0.321 | 0.031 | 0.281 | 0.258 | 0.023 |
| 4th set | 0.794 | 0.849 | -0.055 | 0.453 | 0.474 | -0.021 | 0.341 | 0.375 | -0.034 |
| 5th set | 0.769 | 0.818 | -0.049 | 0.497 | 0.477 | 0.020 | 0.272 | 0.341 | -0.069 |
| 6th set | 0.824 | 0.823 | 0.001 | 0.502 | 0.463 | 0.039 | 0.322 | 0.361 | -0.039 |
| 7th set | 0.750 | 0.691 | 0.059 | 0.411 | 0.396 | 0.015 | 0.339 | 0.295 | 0.044 |
| RMSE |  |  | 0.0416 |  |  | 0.0231 |  |  | 0.0377 |

As expected, the simulation results of bidirectional flow rate are smaller than that of the unidirectional flow rate, and it is in agreement with the experimental ones, see Fig.3a and Tables 2 and 4, which present a quantitative comparison between experimental results and simulation results.

## 5. Conclusion

In this paper, we have experimentally studied the impact of holding umbrella on the uni- and bi-directional flow. Seven sets of experiments have been implemented, in which pedestrians holding umbrellas of different colors are asked to walk clockwise or anti-clockwise in a ring-shaped corridor. The flow rate under the umbrella situation is smaller than that in the normal condition, no matter in uni- or bi-directional flow. In bidirectional flow, pedestrians under the normal situation segregate into two opposite moving streams very quickly, and clockwise/anti-clockwise walking pedestrians are always in the inner/outer ring due to right-walking preference. In experiments under holding umbrella situation, spontaneous lane formation has also occurred, but clockwise/anti-clockwise pedestrians are not always in the inner/outer ring, and there can form more than two lanes in the stable state. Moreover, the merge of lanes have been observed. We have generalized a force-based model by adding the contact force between umbrellas. Simulation results are in agreement with the experimental ones.

In our future work, experiments with more participants should be carried out to study the impact of holding umbrella on pedestrian flow dynamics and to examine the proposed models under high density situations.

**Acknowledgments:**

RJ was supported by the National Basic Research Program of China under Grant No.2012CB725404, the Natural Science Foundation of China under Grant No. 11422221 and 71371175.

**References**

Chattaraj U., Seyfried A., Chakroborty P., 2009. Comparison of pedestrian fundamental diagram across cultures. Advances in Complex Systems **12**, 393-405.

Daamen W. and Hoogendoorn S., 2003. Experimental research of pedestrian walking behavior. Transportation Research Record: Journal of the Transportation Research Board **1828**, 20-30.

Flötteröd G., Lämmel G., 2015. Bidirectional pedestrian fundamental diagram. Transportation Research Part B **71**, 194-212.

Garcimartín A., Zuriguel I., Pastor J.M., Martín-Gómez C., ParisiD.R., 2014. Experimental evidence of the "faster is slower" effect. Transportation Research Procedia **2**, 760-767.

Guo N., Hao Q.Y., Jiang R., Hu M.B., Jia B., 2015. Uni- and Bi-directional pedestrian flow in the view-limited condition: Experiments and modeling. Under review.

Guo R.Y., Wong S.C., Xia Y.H., Huang H.J., Lam W.H.K., Choi K., 2012. Empirical evidence for the look-ahead behavior of pedestrians in bi-directional flows. Chinese physics letters 29, 068901.

Helbing D., Farkas I., Vicsek T., 2000a. Simulating dynamical features of escape panic. Nature **407**, 487-490.

Helbing D., Farkas I.J., Vicsek T., 2000b. Freezing by heating in a driven mesoscopic system. Physical Review Letters **84**, 1240-1243.

Helbing D., 2001. Traffic and related self-driven many-particle systems. Reviews of Modern Physics **73**, 1067-1141.

Helbing D., Johansson A., Al-Abideen H.Z., 2007. Dynamics of crowd disaaters: an empirical study. Physical Review E **75**, 046109.

Helbing D., Johansson A., 2011. Pedestrian, crowd and evacuation dynamics. Extreme Environmental Events, 697-716.

Kretz T., Grünebohm A., Kaufman M., Mazur F., Schreckenberg, M., 2006. Experimental study of pedestrian counterflow in a corridor. Journal of Statistical Mechanics: Theory and Experiment, P10001.

Hoogendoorn S.P., Daamen W., 2005. Pedestrian behavior at bottlenecks. Transportation Science **39**, 147-159.

Hou L., Liu J.G., PanX., Wang B.H., 2014. A social force evacuation model with the leadship effect. Physica A **400**, 93-99.

Moussaïd M., Perozo N., Garnier S., Helbing D., Theraulaz G., 2010. The walking behaviour of pedestrian social groups and its Impact on crowd dynamics. Plos One **5**, e10047.

Moussaïd M., Helbing D., and Theraulaz G., 2011. How simple rules determine pedestrian behavior and crowd disasters. PNAS **108**, 6884–6888.

Moussaïd M., Guillot E.G., Moreau M., Fehrenbach J., Chabiron O., Lemercier S., Pettre J., Appert-Rolland C., Degond P., Theraulaz G., 2012. Traffic instabilities in self-organized pedestrian crowds. Plos Computational Biology **8**,

e1002442.

Nagai R., Fukamachi M., Nagatani T., 2005. Experiment and simulation for counterflow of people going on all fours. Physica A **358**, 516-528.

Navin F.D., Wheeler R.J., 1969. Pedestrian flow characteristics. Traffic Engineering **39**, 31-36.

Nowak S., Schadschneider A., 2012. Quantitative analysis of pedestrian counterflow in a cellular automaton model. Physical Review E 85, 066128.

Schadschneider A., Kirchner A., Nishinari K., 2003. From ant trails to pedestrian dynamics. Applied Bionics and Biomechanics **1**, 11-19.

Seyfried A., Steffen B., Klingsch W., Boltes M., 2005. The fundamental diagram of pedestrian movement revisited. Journal of Statistical Mechanics: Theory and Experiment, P10002.

Sivers I., Köster G., 2015. Dynamic stride length adaptation according to utility and personal space. Transportation Research Part B **74**, 104-117.

Suma Y., Yanagisawa D., Nishinari K., 2012. Anticipation effect in pedestrian dynamics: modeling and experiments. Physica A **391**, 248-263.

Yanagisawa D., Tomoeda A., Nishinari K., 2012. Improvement of pedestrian flow by slow rhythm. Physical Review E **85**, 016111.

Zhang J., Klingsch W., Schadschneider A., Seyfried A., 2012. Ordering in bidirectional pedestrian flows and its influence on the fundamental diagram. Journal of Statistical Mechanics: Theory and Experiment, P02002.

Zhang Q., Seyfried A., 2014. Comparison of intersecting pedestrian flows based on experiments. Physica A **405**, 316-325.